# Synchronising DER inverters to weak grid using Kalman filter and LQR current controller

Phuoc Sang Nguyen, *Student Member, IEEE,* Ghavameddin Nourbakhsh, *Member, IEEE,* and Gerard Ledwich, *Life Fellow, IEEE*

*Abstract*—Grid-following (GFL) inverters are commonly used for integrating renewable energy sources into power grids. However, the dynamic performance of GFL models can be significantly impacted by the Phase-Locked Loop (PLL) in a weak grid, leading to instability due to inaccuracies in grid source phase angle estimation. The proposed method in this manuscript replaces the PLL with an Advanced Angle Estimation based Kalman Filter including a Linear Quadratic Regulator (LQR) controller of the GFL. This method is robust in incorporating grid impedance terms as part of state space models in the Kalman Filter approach to estimate instantaneous phase angle using *α-β* Synchronous Reference Frame equations. The stability performance of the proposed approach is validated through eigenvalue analysis in a two-source case. Additionally, an LQR controller is employed to regulate capacitor voltage, inverter current, and the current at the Point of Common Coupling (PCC). The proposed controller surpasses existing approaches in terms of accuracy and distortion reduction under abrupt grid impedance increases. Moreover, drop compensation is integrated into the Kalman Filter to enhance robustness of the inverter against external oscillation disturbances from a synchronous machine connected to the GFL via the PCC. The results in this paper demonstrate substantial improvement in oscillation damping across a range of frequencies compared with published research works.

## I. Introduction

RECENT years have seen a rapid increase in the incorporation of distributed energy resources (DERs) within electric grids. As a result, GFL are commonly employed in most commercially installed power electronic inverters, known as inverter-based resources [1]. In the literature, a wide variety of grid synchronisation approaches used in GFL inverters are addressed, aiming to accurately track or estimate the angle and frequency of the grid voltage. This ensures system stability and desirable performance [2].

The Conventional Phase-Locked Loop (CPLL) method, commonly used for tracking voltage and frequency to achieve grid synchronisation, is detailed in [3]–[5]. However, integrating the PLL for grid synchronisation in GFL can significantly impact their dynamic performance and stability, mainly in cases with high grid impedance [6]–[8]. Researchers have found that PLL controls can cause low-frequency instability in high grid impedance conditions by demonstrating a negative

Phuoc Sang Nguyen, Ghavameddin Nourbakhsh, and Gerard Ledwich are with the School of Electrical Engineering and Robotics, Queensland University of Technology, Brisbane, QLD 4000, Australia incremental resistance term in d-q impedance-based analysis, as revealed through the generalised Nyquist criterion [9].

Various strategies have been suggested to address instability issues related to grid synchronisation caused by PLL dynamics, with the goal of enhancing overall system performance. A suggested approach involves the development of a damping controller integrated with a band-pass filter (BPF) [10], which mitigates negative resistance from the PLL by regulating of the BPF to a particular frequency, thereby improve stability. Another study reported in [11] introduces a feedforward loop linking the PLL to the current controller to achieve symmetrical dynamics in the d-q frame, thereby enhancing the frequency couplings affected by the PLL. Additionally, a conventional virtual impedance model introduced in [12], known as CVI-PLL, integrates with the PLL to mitigate destabilising effects of the self-synchronisation loop. This model allows the inverter to synchronise virtually with a point that exhibits greater grid robustness. Moreover, a modified version of the virtual impedance method (developed by same authors in this paper), termed MVI-PLL proposed in [13], which incorporates a line drop compensation into the virtual impedance model, enhancing performance across a broader grid strengths. However, single-state feedback from PI controllers shows limitations in weak connections and results in longer convergence times.

Alternatively, the Kalman Filter (KF) is a widely used strategy for grid synchronisation, as discussed in [14] & [15], where the authors highlight that leveraging system measurements and dynamics allows the KF to effectively reduce disturbances, providing precise and rapid estimation of instantaneous grid angle, even in the presence of Gaussian noise. Furthermore, [16] has introduced a Kalman filter method referred here as Conventional Angle Estimation based Kalman Filter (CAEKF), which applies a linear parametric model of grid voltage in *α-β* frame to estimate frequency and phase angle. However, this method primarily concentrates on measurements at the grid terminals and neglects grid impedance. In practice, measurements are normally acquired from the point of common coupling (PCC) on the inverter side. While grid impedance is crucial due to its significant impact on system stability.

Designing a current controller for an LCL-filter of a GFL inverter is also a key factor in improving computational time, also ensuring stability performance. Typically, the most popular current control methods for LCL-filters have been established in a single-input–single-output controller, with the proportional–integral (PI) control structure being one of the most commonly employed method [17]. However, a disadvantage of this controller is that the transfer function between the inverter and grid side currents present an underdamped resonance, leading to undesired oscillations in the grid current [18]. An additional challenge involves in identifying the suitable controller gains to stabilise the system and ensure effective performance in steady-state and transient conditions when using a multivariable design method [19]. Numerous studies on the selection of gains for a full-state feedback controller enhanced with multiple control terms are discussed in [19]–[22] using various methodologies. In [21], [22], a simple approach utilising the pole placement technique is presented for selecting system gains. However, a complicated system results in a higher number of gains, which reduces the efficiency of this method due to their time-consuming implementation [20]. In contrast, authors in [19] have successfully applied a linear quadratic regulator (LQR) for an LCL filter grid connected inverter. This approach utilised to systematically select feedback gains by minimising the discrete cost function, ensuring stability and robustness requirements of the system.

The literature review highlights that existing methods largely neglect the impact of grid impedance in their controllers. Furthermore, single-state feedback from PI current controllers can be poorly damped and poses challenges in identifying appropriate controller gains for multivariable systems [19]–[22]. This paper introduces the Advanced Angle Estimation based Kalman Filter using an LQR current controller (AAEKF-LQR), an enhanced version of CAEKF. This approach integrates grid impedance terms into the KF method to estimate instantaneous phase angle. Additionally, the LQR current controller is applied to enhance voltage measurement accuracy and dynamic system performance. Numerical simulations of AAEKF-LQR are conducted to evaluate and validate the stability and dynamic performance of a dual-source system, considering variations in key influencing parameter values. Furthermore, under conditions of high grid impedance, the stability performance of the five noted methods outlined in Table I are assessed and compared. The assessments are done through discrete-domain simulations conducted in the Matlab/Simulink environment. Fig. 1 shows the integration of the AAEKF-LQR approach into a GFL.

TABLE I
FIVE GRID SYNCHRONISATION METHODS OF GFL

| Abbreviation | Description |
| --- | --- |
| CPLL | Conventional Phase-locked loop |
| CVI-PLL | Conventional Virtual Impedance - Phase-locked loop |
| MVI-PLL | Modified Virtual Impedance - Phase-locked loop |
| CAEKF | Conventional Angle Estimation based Kalman Filter |
| AAEKF-LQR | Advanced Angle Estimation based Kalman Filter - Linear Quadratic Regulator Current controller |

The primary contributions of this paper include:

1) Introducing a model-based AAEKF to replace the model-free PLL in GFL systems, incorporating line drop compensation to estimate the instantaneous grid phase angle under high grid impedance conditions.
2) Implementing an LQR controller to replace the PI-based inner current controller, enabling regulation of

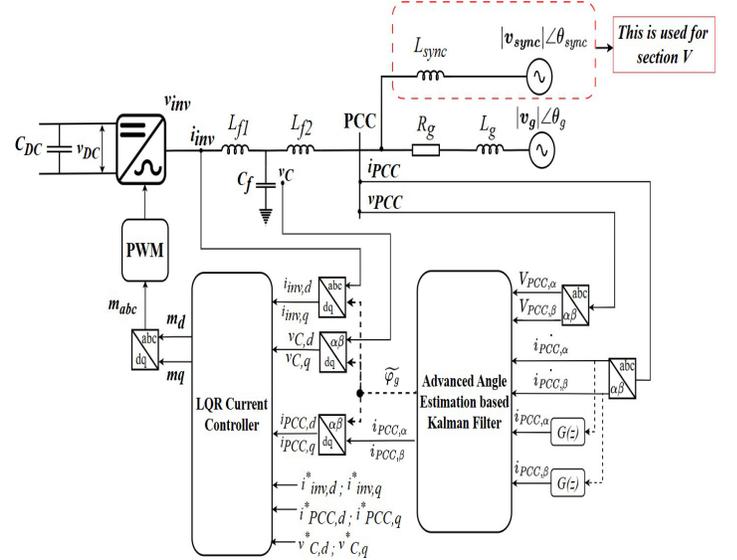

Fig. 1. Grid-following inverter model with AAEKF-LQR approach.

three-state feedback to minimise variations and mitigate system oscillations.
3) Demonstrating the robust stability performance of the proposed AAEKF method for different grid impedances with sudden increase and grid impedance model errors.
4) Highlighting the robustness of the proposed method in addressing external oscillation disturbances introduced by a synchronous machine connected to the GFL supported network.

## II. GRID-FOLLOWING INVERTER MODEL WITH AAEKF METHOD DESCRIPTION

The proposed approach aims to estimate the instantaneous grid phase angle $\varphi_g$. In this section, the electrical system's Synchronous Reference Frame (SRF) equations on the AC side are described in the $\alpha$ and $\beta$ frames. The grid voltage in the $\alpha$-$\beta$ frames is given as in (1) and (2).

$$\frac{dV_{g,\alpha}}{dt} = -\omega_g V_{g,\beta}(t) \quad (1)$$

$$\frac{dV_{g,\beta}}{dt} = \omega_g V_{g,\alpha}(t) \quad (2)$$

where $V_{g,\alpha}(t) = |V_g|\cos[\varphi_g(t)]$, $V_{g,\beta}(t) = |V_g|\sin[\varphi_g(t)]$, $\varphi_g(t) = \omega_g t + \theta_g$, $\theta_g$ is grid angle in the top right of Fig. 1. Furthermore, $\omega_g = 2\pi f_g$ is the grid frequency. Additionally, the PCC voltage through grid impedance $R_g$ & $L_g$ in the $\alpha$-$\beta$ frame is presented as:

$$V_{PCC,\alpha}(t) = V_{g,\alpha}(t) + R_g i_{PCC,\alpha}(t) - \omega_g L_g i_{PCC,\beta}(t) \quad (3)$$

$$V_{PCC,\beta}(t) = V_{g,\beta}(t) + R_g i_{PCC,\beta}(t) + \omega_g L_g i_{PCC,\alpha}(t) \quad (4)$$

The grid voltage in $\alpha$-$\beta$ frame is selected as the system state variables, $x_{KF}(t) = [V_{g,\alpha}(t), V_{g,\beta}(t)]^T$. The input is $u_{KF}(t) = [i_{PCC,\alpha}(t), i_{PCC,\beta}(t)]^T$. By considering, $t = kT_s$, with $T_s = 10^{-4}$s being the sampling period and $k$ is the sampling instant. The Kalman Filter operates as a state-optimal estimator, requiring a state space model for implementation.

The state space equations derived from the previously noted as in (5), where $y_{KF}(t)$ is the PCC voltage in $\alpha$-$\beta$ frame.

$$\begin{aligned} \dot{x}_{KF}(t) &= \begin{bmatrix} 0 & -\omega \\ \omega & 0 \end{bmatrix} x_{KF}(t) + \begin{bmatrix} R_g & -\omega_g L_g \\ \omega_g L_g & R_g \end{bmatrix} u_{KF}(t) \\ y_{KF}(t) &= \begin{bmatrix} 1 & 0 \\ 0 & 1 \end{bmatrix} x_{KF}(t) \end{aligned} \quad (5)$$

In order to implement the KF digitally, the state equations in the continuous time need to be discretised. The discrete state-space equations are described as in (6).

$$\begin{aligned} x_{KF}(k+1) &= A_d x_{KF}(k) + v(k) \\ y_{KF}(k) &= C_d x_{KF}(k) + D_d u(k) + w(k) \end{aligned} \quad (6)$$

where $w(k)$ and $v(k)$ represent the process noise and the measurement noise, which are used to define the process covariance matrix $Q_{KF}$ and the measurement covariance matrix $R_{KF}$ in (7).

$$Q_{KF} = E\left[ v(k) v^T(k) \right] = q_{kf} I_2; \quad R_{KF} = E\left[ w(k) w^T(k) \right] \quad (7)$$

where $I_2$ is the $2 \times 2$ Identity matrix and $q_{kf}$ is a scalar value. The discrete system matrix and input matrix $A_d$, $C_d$ and $D_d$ can be derived as in [23]. The implementation of the KF algorithm is based on [24]. However, in this proposed approach, the KF estimator incorporates line drop compensation, resulting in the estimated states $\hat{x}_{KF}(k)$ being computed as follows.

$$\hat{x}_{KF}(k) = A_d \hat{x}_{KF}(k-1) + K_{KF}(k)[y_{KF}(k) - \hat{y}_{KF}(k)] \quad (8)$$

$$\hat{y}_{KF}(k) = C_d \hat{x}_{KF}(k \mid k-1) + D_d u_{KF}(k-1) \quad (9)$$

where $K_{KF}(k)$ is the Kalman gain.

By applying the KF procedure above, the grid voltage in $\alpha$-$\beta$ frame are estimated. From these results, the estimated phase angle is computed as in (10).

$$\hat{\varphi}(k) = \tan^{-1}\left\{ \frac{\hat{V}_{g,\beta}(k)}{\hat{V}_{g,\alpha}(k)} \right\} \quad (10)$$

where $\hat{V}_{g,\alpha}(k) = |\hat{V}|\cos[\hat{\varphi}(k)]$, $\hat{V}_{g,\beta}(k) = |\hat{V}|\sin[\hat{\varphi}(k)]$ and $\varphi_g(k) = \omega_g k T_s + \theta_g$ is the estimated instantaneous grid phase angle.

Algorithm 1 is showing the procedure of AAEKF.

---

**Algorithm 1**: AAEKF algorithm for GFL inverter.

---

**Input**: a measurement matrix $V_{PCC}$ and an input matrix $i_{PCC}$ in $\alpha$-$\beta$ frame, measurement noise covariance $R_{KF}$, process noise covariance $Q_{KF}$, and grid impedance $R_g + j\omega_g L_g$.
**Output**: estimated instantaneous phase angle $\hat{\varphi}$.

---

1: State variables $\hat{x}(t) = \left[ \hat{V}_{g,\alpha}, \hat{V}_{g,\beta} \right]^T$
   Input $u(t) = [i_{PCC,\alpha}, i_{PCC,\beta}]^T$
   $T_s = 10^{-4}$s.
2: Establish state space model $ABCD$ matrix in continuous-time domain as in (5).
3: Convert to discrete state space model $A_d, B_d, C_d$ and $D_d$ matrix from (6).
4: Start $k=0$. Initialise estimated state $\hat{x}_{KF}(0)$ and error covariance matrix $P(0)$ as the flat start.
5: $k = k+1$
6: Apply KF into discrete state space model.
7: *Prediction* :
   Calculate $\hat{x}_{KF}(k \mid k-1)$ and $P(k \mid k-1)$ from [24].
8: *Estimation* :
9: Update Kalman gain $K_{KF}(k)$, estimated state $\hat{x}_{KF}(k)$. Compute instantaneous phase angle.
   $\hat{\varphi}_g(k) = \tan^{-1}\left( \frac{\hat{V}_{g,\beta}(k)}{\hat{V}_{g,\alpha}(k)} \right)$.
10: Move to step 5, continue until simulation is completed.
11: **return** $\hat{\varphi}_g(k)$.

---

## III. LQR Current Controller's design

This section focuses on designing an LQR controller to regulate the PCC current, inverter current and capacitor voltage. The current controller utilises the d-q frame for implementation, enabling efficient operation within the system. The advantage of using the d-q frame lies in its ability to simplify control: when the system is stable or balanced in three-phase operation, the d-q frame values remain constant, making them easier to manage and control. The equations that govern the behaviour of the electrical system in the Synchronous Reference Frame (SRF) on the AC-side of the GFL model are written in the d-q frame, as given by (11).

$$\mathbf{x}(k) = x_d(k) + j x_q(k) \quad (11)$$

From Fig. 1, by considering the AC side from PCC to inverter, the corresponding equations describing this model are given in (12), (13) and (14):

$$\frac{d\mathbf{i_{inv}}}{dt} = \frac{1}{L_{f1}}(\mathbf{v_{inv}} - \mathbf{v_C} - j\omega_g L_{f1} \mathbf{i_{inv}}) \quad (12)$$

$$\frac{d\mathbf{i_{PCC}}}{dt} = \frac{1}{L_{f2}}(\mathbf{v_C} - \mathbf{v_{PCC}} - j\omega_g L_{f2} \mathbf{i_{PCC}}) \quad (13)$$

$$\frac{d\mathbf{v_C}}{dt} = \frac{1}{C_f}(\mathbf{i_{inv}} - \mathbf{i_{PCC}} - j\omega_g C_f \mathbf{v_C}) \quad (14)$$

The inverter current, PCC current and filter capacitor voltage in d-q frame are selected as the system state variables be $x_{LQR}(t) = [i_{inv,d}(t), i_{inv,q}(t), i_{PCC,d}(t), i_{PCC,q}(t), v_{C,d}(t), v_{C,q}(t)]^T$. The inverter voltage is selected as input $u_{LQR}(t) = [v_{inv,d}(t), v_{inv,q}(t)]^T$. The state space equations derived from the previously noted equations are outlined in (15).

$$\dot{x}_{LQR}(t) = A_{LQR} x_{LQR}(t) + B_{LQR,1} u_{LQR}(t) + B_{LQR,2} \begin{bmatrix} v_{PCC_d}(t) \\ v_{PCC_q}(t) \end{bmatrix} \quad (15)$$

The system matrix $A_{LQR}$, and input matrices $B_{LQR,1}$, $B_{LQR,2}$ in the continuous-time domain are given in (16) and (17).

$$A_{LQR} = \begin{bmatrix} 0 & \omega_g & 0 & 0 & -1/L_{f1} & 0 \\ -\omega_g & 0 & 0 & 0 & 0 & -1/L_{f1} \\ 0 & 0 & 0 & \omega_g & 1/L_{f2} & 0 \\ 0 & 0 & -\omega_g & 0 & 0 & 1/L_{f2} \\ 1/C_f & 0 & -1/C_f & 0 & 0 & \omega_g \\ 0 & 1/C_f & 0 & -1/C_f & -\omega_g & 0 \end{bmatrix} \quad (16)$$

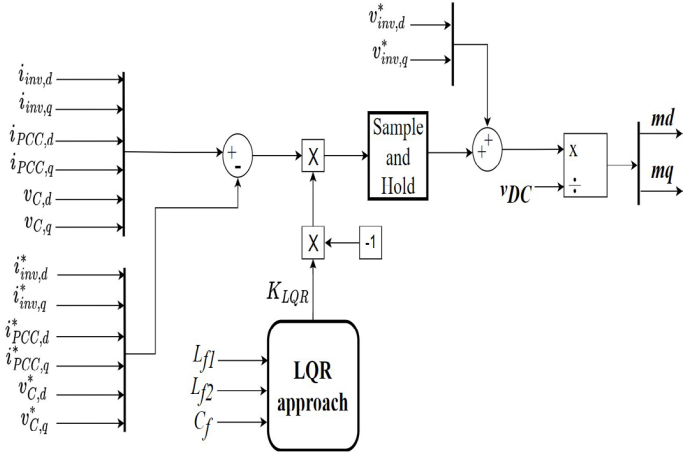

Fig. 2. A LQR current controller.

$$B_{LQR,1} = \begin{bmatrix} 1/L_{f1} & 0 \\ 0 & 1/L_{f1} \\ 0 & 0 \\ 0 & 0 \\ 0 & 0 \\ 0 & 0 \end{bmatrix}; \quad B_{LQR,2} = \begin{bmatrix} 0 & 0 \\ 0 & 0 \\ -1/L_{f2} & 0 \\ 0 & -1/L_{f2} \\ 0 & 0 \\ 0 & 0 \end{bmatrix} \quad (17)$$

This section focuses on controlling the PCC current, inverter current, and capacitor voltage to achieve the desired reference value. To determine the reference values for the states, the system's equilibrium point $\overline{x_{LQR}}$ is established. This involves setting the time derivatives of the state variables to zero, implying $\dot{\overline{x_{LQR}}} = 0$. Solving these equations yields the values of the state variables at the equilibrium point $\overline{x_{LQR}}$. This autonomous system does not consider disturbance and a model of line impedance.

An optimal controller is employed to achieve the desired performance by considering a mathematical function that penalises deviations from the desired performance, known as LQR optimal control. This approach aids in achieving optimal pole locations and minimises the cost of the control system. The cost function is defined in (18). Here, $R_{LQR}$ represents the penalty on control actions, while $Q_{LQR}$ represents penalties on deviations in the state variables.

$$J = \int_0^\infty \left( x_{LQR}^T Q_{LQR} x_{LQR} + u_{LQR}^T R_{LQR} u_{LQR} \right) dt, \quad (18)$$
$$Q_{LQR} \geq 0, \quad R_{LQR} > 0$$

Similar to the Kalman Filter, the state equations in continuous time need to be discretised. The sampling time is $10^{-4}$s. The discrete system matrix $A_{LQR\_d}$ and input matrix $B_{LQR,1\_d}$ can be derived as follows by [23]. Moreover, the matrices of $Q_{LQR}$ and $R_{LQR}$ are selected as in (19). Where $r_{lqr}$ is a scalar value.

$$Q_{LQR} = \mathrm{diag}\left( [q_{lqr,1}, q_{lqr,1}, q_{lqr,2}, q_{lqr,2}, q_{lqr,3}, q_{lqr,3}]^T \right); \quad (19)$$
$$R_{LQR} = r_{lqr} I_2$$

The controller gain $K_{LQR}$ in steady state is computed by $K_{LQR} = R_{LQR}^{-1} B_{LQR,1}^T S$, where $S$ is a positive semi-definite symetric matrix of the Algebraic Riccati Equation [25]. In this context, the design of an LQR controller is aimed to regulate the six states in the d-q frame, including the inverter current, PCC current, and capacitor voltage. The optimal controller is written as in (20). The LQR Current Controller block is implemented in Fig. 2.

$$u_{LQR}^*(k) = -K_{LQR}\left[ x_{LQR}(k) - \overline{x_{LQR}} \right] \quad (20)$$

IV. SYSTEM SIMULATION AND ANALYSIS

A. Case study

TABLE II
PARAMETERS OF CONTROL AND ELECTRICAL SYSTEMS [13].

| Parameter | Value | Parameter | Value |
|---|---|---|---|
| Rated power | 110 KVA | Filter capacitance $C_f$ | 100 $\mu$F |
| Rated voltage | 415 V | Filter inductance $L_{f1} = L_{f2}$ | 500 $\mu$H |
| Grid voltage $v_g$ | 1 pu | Reference inverter current $i^*_{inv}$ | $1\angle 30°$ pu |
| Grid frequency $\omega_g$ | $2\pi \times 50$ rad/s | DC source voltage $v_{DC}$ | 2.5 pu |

This study employs numerical simulations to validate the stability of the proposed AAEKF-LQR approach by analysing the eigenvalues of $\vec{A}_{error}$ matrix derived from the state space model of prediction and output error within the Kalman filter, as outlined in [16]. The $\vec{A}_{error}$ matrix can be expressed as (21).

$$\vec{A}_{error}(k) = A_d\left[ I_2 - K_{KF}(k) C_d \right] \quad (21)$$

Furthermore, this section provides a comparative analysis of five methodologies: CPLL, CVI-PLL, MVI-PLL, CAEKF, and AAEKF-LQR, assessing their accuracy and dynamic performance within a two-source system framework. MATLAB/Simulink simulations of these five approaches are conducted, with the parameters for both the electrical system and control strategies detailed in Table II. To ensure consistency, all parameters are standardised to per unit. The implementation procedures for CPLL, CVI-PLL, and MVI-PLL follow the methodologies described in [3], [12] & [13], respectively. In the system diagram illustrated in Fig. 1, the AAEKF block is replaced by the PLL block. Additionally, the proportional and integral gains of the PI controller within the PLL for the three specified methods are selected, as outlined in Table II, with $k_p = k_i = 5$, enabling these methods to achieve similar performance comparable to that observed in other studies in [3], [12] & [13].

This work assumes a constant DC link voltage. Two primary methods exist for mitigating harmonics in the DC link output voltage. The first involves employing a relatively large capacitor to suppress voltage ripples; however, this approach reduces the dynamic response of the PWM, slows the outer control loop, and results in delayed reference tracking [26] & [27]. The second method utilises a nonlinear control strategy to minimise small disturbances, enabling the use of a smaller DC link capacitor [28]–[31]. Moreover, this study also assumes the grid impedance to be known. Additional work has shown that the grid impedance is identifiable. However, these topics are beyond the scope of the current analysis.

B. Model Validation by the proposed approach AAEKF-LQR

This section utilises an actual grid impedance value of $2\angle 70°$ pu for model validation and analysis for distribution case studies.

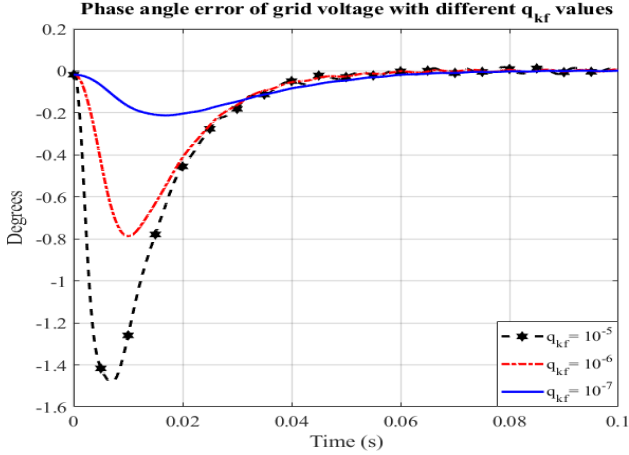

Fig. 3. Phase angle error of grid voltage with different $q_{kf}$ values.

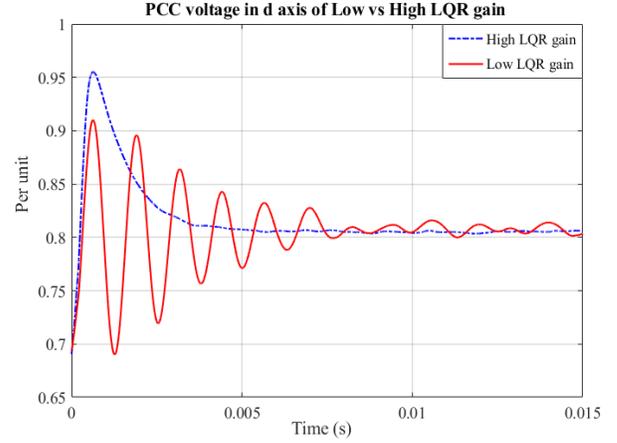

Fig. 4. Per unit PCC voltage in d frame of Low vs High LQR gain.

*1) Evaluate Advanced Angle estimation based Kalman Filter approach:* In this part, the effectiveness of the AAEKF method is validated. This validation involves varying the covariance process noise matrix $Q_{KF}$. Additionally, to minimise the influence of the LQR controller, small values for the $K_{LQR}$ gain are utilised. This approach enables a more effectively isolated assessment of the impact of angle estimation using the KF. For this example, the matrices $Q_{LQR}$ and $R_{LQR}$ are selected as $Q_{LQR} = diag([10, 10, 10, 10, 10, 10]^T)$ and $R_{LQR} = 10^{-2}I_2$, respectively, resulting in the following gain $K_{LQR}$ as shown in (22). This $K_{LQR}$ values demonstrate symmetry and decoupling between the d and q axis, indicating that control of the d axis is largely dependent on d axis measurements.

$$K_{LQR} = \begin{bmatrix} 0.426 & 0.007 & 0.002 & 0.001 & 0.034 & 0.001 \\ -0.007 & 0.426 & -0.001 & 0.002 & -0.001 & 0.034 \end{bmatrix} \quad (22)$$

From (22), the $A_{error}$ matrix, which comprises state-space difference equations governing the state prediction and output error dynamics of the KF, is examined for stability. Numerical simulations are conducted using three distinct $q_{kf}$ values of the $Q_{KF}$ matrix: $10^{-5}$, $10^{-6}$, and $10^{-7}$, all with same $R_{KF} = I_2$. In discrete-time systems, stability is assured when the eigenvalues $\lambda$ of the $A_{error}$ matrix satisfy the condition $|\lambda| \leq 1$. The simulation results indicate that the steady-state Kalman Filter eigenvalues, although exceeding 0.99, consistently remain below 1 for these three $q_{kf}$ values. This confirms the system's stability and allows for further analysis of its behaviour.

Fig. 3 illustrates that the instantaneous phase angle error between the estimated and actual grid voltage angles trends towards zero for all three $q_{kf}$ values. Higher $q_{kf}$ values result in a faster reduction of the estimated phase angle error but also display higher initial angle error until convergence. Consequently, a $q_{kf}$ value of $10^{-6}$ is chosen for subsequent simulations.

*2) Evaluate LQR current controller:* To evaluate the impact of the LQR controller on the system, the accuracy and dynamic performance of the PCC voltage in the d-q frame are compared between low and high $K_{LQR}$ gains. The low $K_{LQR}$ gain values have been mentioned previously. For the high $K_{LQR}$ gain, the $Q_{LQR}$ and $R_{LQR}$ matrices are selected as $Q_{LQR} = diag([10^3, 10^3, 10^3, 10^3, 10, 10]^T)$ and $R_{LQR} = 10^{-2}I_2$, respectively. The resulting gain $K_{LQR}$ is shown below:

$$K_{LQR} = \begin{bmatrix} 5.292 & 0.085 & 0.317 & 0.003 & 0.943 & 0.015 \\ -0.085 & 5.292 & -0.003 & 0.317 & -0.015 & 0.943 \end{bmatrix} \quad (23)$$

This $K_{LQR}$ gain meets the expectation of maintaining symmetry and decoupling between the d and q axes similar to the low $K_{LQR}$ gain. Furthermore, in this high $K_{LQR}$ gain scenario, higher penalty values were selected for the inverter current and PCC current states compared to the capacitor voltage state. Fundamentally aimed at controlling current, this approach incorporates voltage feedback to enhance stabilisation purposes. As validated in multiple simulations, inverter and PCC current states exhibited higher errors and longer convergence times than the capacitor voltage state. Consequently, higher penalty values in the $Q_{LQR}$ matrix are necessary to enhance performance.

Fig. 4 demonstrates that the PCC voltage in the d frame stabilises more quickly with a high $K_{LQR}$ gain, achieving stability in 0.006 seconds, while the PCC voltage with a low $K_{LQR}$ gain continues to oscillate at 0.015 seconds. Additionally, the low $K_{LQR}$ gain permits significant oscillations in the waveform, indicating less effective control performance compared to the high $K_{LQR}$ gain. Thus, a higher $K_{LQR}$ gain improves system performance and is used for further analysis and simulations.

*3) Stability Analysis of AAEKF-LQR Method Under Grid Impedance Model Errors:* This section examines the stability performance of the system when grid impedance model errors are input into the AAEKF method and the LQR current controller. Fig. 5 illustrates the phase angle error of the grid voltage for different percentages of the input error of the grid impedance in the AAEKF-LQR method. The results indicate that generally the smaller grid impedance model errors result in reduced phase angle errors in the grid voltage. Additionally, the AAEKF-LQR method demonstrates robust control and system stabilisation capabilities with up to ±20% error in grid impedance model. However, significant phase angle errors may result in instability and increased reference tracking errors related to capacitor voltage, PCC and inverter current.

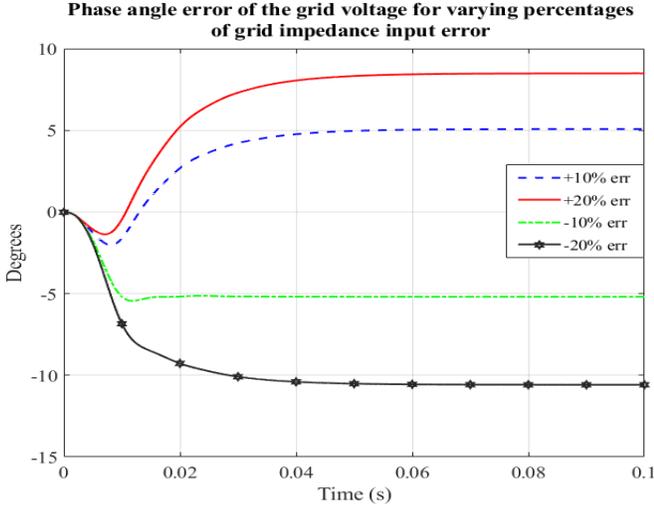

Fig. 5. Instantaneous phase angle error of the grid voltage for varying percentages of grid impedance model error in the AAEKF-LQR method.

*C. Analysis of Proposed Method AAEKF-LQR versus other four methods under Sudden Grid Impedance Increases During the Simulation Process*

In this section, discrete-domain simulations of five methods are utilised to evaluate and compare their efficacy in terms of dynamic performance and stability characteristics under conditions of increases in grid impedance. This evaluation aims to demonstrate the impact of varying grid strengths on the system's recovery, accuracy, and convergence time, offering valuable insights into the system's response under different grid conditions. Fig. 6 presents the PCC voltage of CPLL, CVI-PLL, MVI-PLL, CAEKF, and AAEKF-LQR under varying grid impedance magnitudes while maintaining a grid impedance angle of 70 degrees. The grid impedance magnitude is sequentially increased to 0.6, 1.2, 1.9, and 2.2 pu at intervals of 0.08 seconds. The CPLL and CAEKF present considerable voltage oscillations and demonstrate limited capability in stabilising when the Short circuit ratio $SCR = \frac{1}{|Z|}$ below 0.833. This limitation is discussed in [13], where it is noted that neglecting terms involving $(R_g + j\omega_g L_g) \times i_{PCC}$ in the synchronisation voltage equation can lead to system destabilisation, particularly under high magnitudes of grid impedance. In contrast, the results demonstrate that CVI-PLL, MVI-PLL and CAEKF exhibit effective voltage recovery capabilities and stability in response to sudden changes in grid impedance, highlighting the critical role of line drop compensation in improving system stability.

On the other hand, the PCC voltage waveform of the CVI-PLL exhibits abnormalities, with noticeable clipping at the waveform peak starting from $|Z| = 1.9$ pu, while the MVI-PLL shows clipping beginning at $|Z| = 2.2$ pu. This occurs because, as the inverter voltage approaches saturation at 1 pu, the PI current controller of CVI-PLL and MVI-PLL, which regulates a single state inverter current, becomes highly sensitive. In contrast, the AAEKF-LQR displays enhanced accuracy and superior performance under the same conditions due to the LQR controller's ability to regulate three states and maintain desired values. Through minimising a quadratic cost function including state errors and control effort, LQR controller effectively enhances control performance, notably in mitigating overshoot and steady-state error.

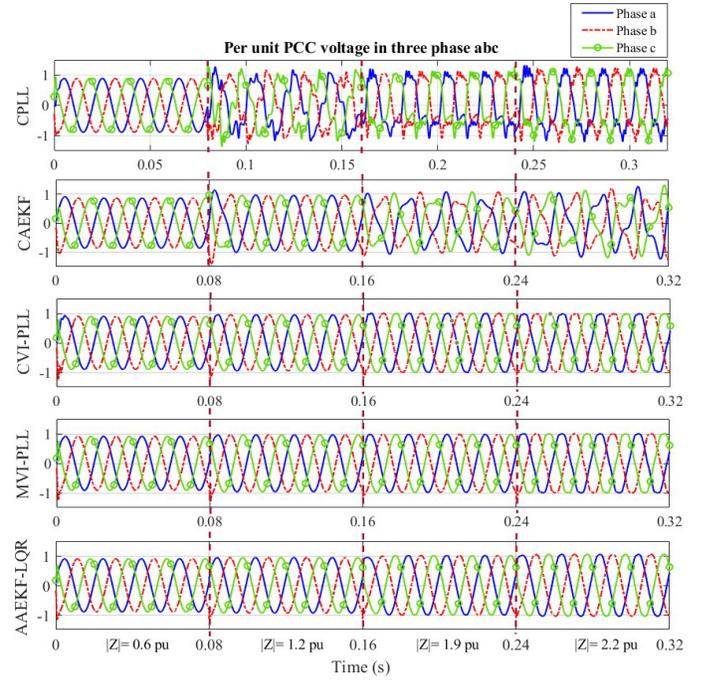

Fig. 6. Per unit PCC voltage of CPLL, CVI-PLL, MVI-PLL, CAEKF, and AAEKF-LQR under varying grid impedance magnitudes.

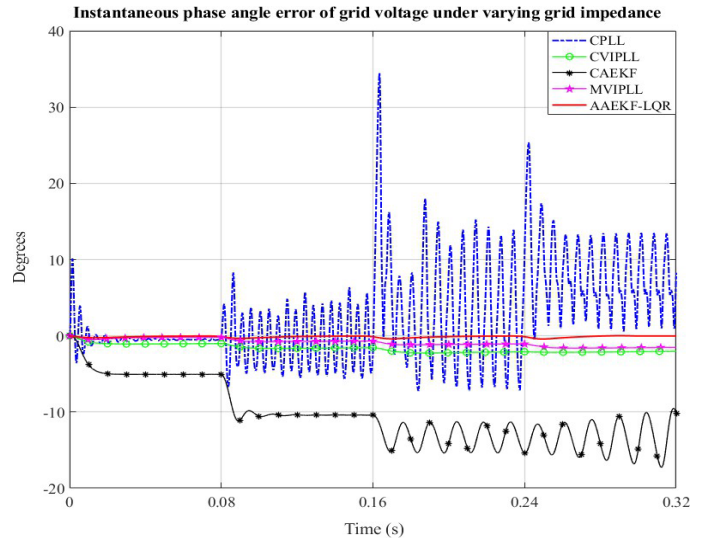

Fig. 7. Instantaneous phase angle error of five methods under varying grid impedance magnitudes situation.

Fig. 7 displays the instantaneous phase angle error of the five methods under varying grid impedance magnitudes situation. The graph indicates that the instantaneous phase angle error of the AAEKF-LQR is closest to zero compared to other methods once the system has converged. The reason is that the KF reduces mean squared error by consistently updating state estimates with each new measurement while maintaining a small error covariance matrix. By considering process and measurement noise and determining an optimal gain for each update, the filter efficiently balances prediction and correction steps.

## V. Robustness of MVI-PLL and AAEKF-LQR against the impact of external disturbances on the system

This section evaluates the impact of the inverter system and the robustness of the two advanced methods, MVI-PLL and AAEKF-LQR demonstrated effective performance under high grid impedance conditions, against external disturbances. These disturbances are introduced into the system and compared with scenarios where the inverter side is disconnected from the GFL model. The external disturbance originates from a synchronous machine connected to the GFL model at the PCC via a purely inductive transmission line. In this study, the parameters for both the electrical system and control are identical to those listed in Table II. The parameters specific to the synchronous machine are detailed in Table III.

In order to evaluate the robustness of two advanced above methods, this case study assumes that both methods do not know any information about the synchronous machine. This scenario examines external disturbance with oscillation frequency range of 3 Hz to 15 Hz, by varying the inertia coefficient of synchronous machine.

Fig. 8 illustrates the load angle of the synchronous machine at an oscillation frequency of approximately 8 Hz. When the system is simulated with only the grid side connected to the synchronous machine and the inverter disconnected, the load angle is affected by the 8 Hz oscillation and continues to oscillate for more than 1.2 seconds. Conversely, when the inverter side is connected and the CPLL technique is employed, the low frequency component converges in approximately 1.5 seconds, but higher frequency components still persist due to the lack of a line drop compensation term and the impact of oscillations on the PLL model's frequency locking. In contrast, the implementation of MVI-PLL and AAEKF-LQR methods effectively manages oscillation effects and enhances convergence. The graph indicates that the load angle with AAEKF-LQR reduces oscillation and converges faster compared to MVI-PLL, with convergence times of 0.8 seconds versus 1.1 seconds, respectively.

TABLE IV
TIME CONSTANT OF AMPLITUDE DECAY.

| Oscillation frequency | GFL without inverter | CVI-PLL | MVI-PLL | AAEKF-LQR |
|---|---|---|---|---|
| 3 Hz | 2.18 s | 2.42 s | 0.55 s | 0.21 s |
| 8 Hz | 0.563 s | 0.588 s | 0.192 s | 0.088 s |
| 15 Hz | 0.126 s | 0.18 s | 0.109 s | 0.118 s |

Additional simulations with oscillation frequencies ranging from 3 Hz to 15 Hz were conducted, and the resulting Table IV records the time constant of amplitude decay to 37% of the initial amplitude for the four cases. The outcomes confirm that both the AAEKF-LQR and MVI-PLL methods demonstrate smaller time constants and faster convergence times compared to those in other cases. Moreover, when the inverter has a similar rating to the synchronous machine, both synchronisation methods have successfully provided rapid oscillation damping. This illustrates the crucial role of line drop compensation, not only in enhancing system stability but also in improving system damping. Additionally, the Kalman Filter presents advantages by prioritising the 50 Hz frequency and treating frequency fluctuations as system process noise. Through suitably identifying the covariance matrix for process noise and employing filtering, the KF adapts to these fluctuations, adjusting its model parameters to have low response to frequencies other than 50 Hz and maintain focus on them exclusively. Furthermore, three state feedbacks are utilised in the LQR controller, with the primary goal of minimising variations in PCC current, inverter current, and capacitor voltage from 50 Hz reference, thereby mitigating the oscillation influence of other synchronous machines on the system.

## VI. Conclusion

This paper introduces the Advanced Angle Estimation based Kalman Filter using an LQR current controller, which integrates grid impedance terms into the KF to enhance voltage measurement accuracy and system stability under high grid impedance conditions. Numerical simulations of AAEKF-LQR, including eigenvalue analysis with varied process covariance matrix $Q_{KF}$ values, validate its performance. Applying the LQR current controller improves system voltage profile and accelerates convergence. Furthermore, the AAEKF-LQR approach effectively ensures robust control and stabilises the system despite grid impedance model error of up to 20%. Moreover, stability comparisons among the five methods CPLL, CVI-PLL, MVI-PLL, CAEKF, and AAEKF-LQR under sudden increases in grid impedance show that AAEKF-LQR achieves faster convergence and superior accuracy with reduced distortion. By incorporating grid impedance in the state space model, AAEKF-LQR provides the most accurate estimated phase angle compared to other methods. Additionally, the robustness of the proposed method against external oscillation disturbances from a synchronous machine connected to the grid via PCC is demonstrated. Both AAEKF-LQR and MVI-PLL methods enhance system damping across an oscillation frequency range of 3 Hz to 15 Hz compared to other approaches. These findings emphasise the essential function of the line drop compensation term in KF in prioritising the 50 Hz frequency, treating frequency variations as system noise, and adjusting model parameters to focus exclusively on maintaining stability at that frequency.

TABLE III
PARAMETERS OF SYNCHRONOUS MACHINE.

| Parameter | Value | Parameter | Value |
|---|---|---|---|
| Synchronous voltage $v_{sync}$ | 1 pu | Synchronous inductance $L_{sync}$ | 300 $\mu$H |
| Synchronous frequency | $2\pi \times 50$ rad/s | Grid impedance | $2\angle 70°$ pu |
| Field Resistance | $5.8 \times 10^{-2}$ pu | Stator Resistance | $2.9 \times 10^{-2}$ pu |

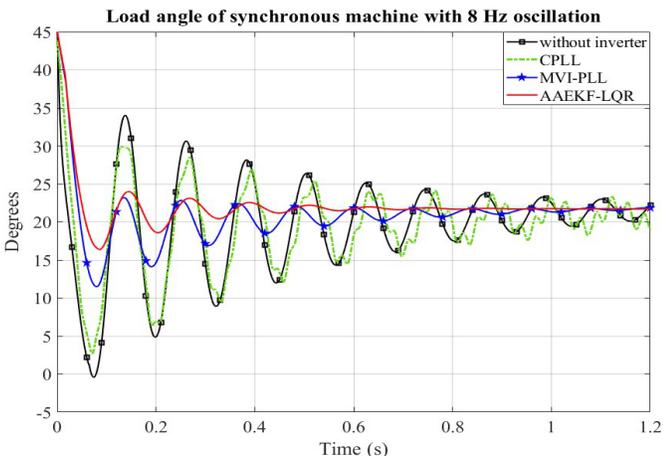

Fig. 8. Load angle of the synchronous machine at an oscillation frequency of approximately 8 Hz in 1.2 seconds.

REFERENCES

[1] F. Blaabjerg, R. Teodorescu, M. Liserre, and A. V. Timbus, "Overview of control and grid synchronization for distributed power generation systems," *IEEE Transactions on industrial electronics*, vol. 53, no. 5, pp. 1398–1409, 2006, ISSN: 0278-0046.
[2] A. Sajadi, J. A. Rañola, R. W. Kenyon, B.-M. Hodge, and B. Mather, "Dynamics and stability of power systems with high shares of grid-following inverter-based resources: A tutorial," *IEEE Access*, 2023, ISSN: 2169-3536.
[3] D. Pattabiraman, R. H. Lasseter, and T. M. Jahns, "Impact of phase-locked loop control on the stability of a high inverter penetration power system," in *2019 IEEE power & energy society general meeting (PESGM)*, IEEE, pp. 1–5, ISBN: 1728119812.
[4] V. Kaura and V. Blasko, "Operation of a phase locked loop system under distorted utility conditions," *IEEE Transactions on Industry applications*, vol. 33, no. 1, pp. 58–63, 1997, ISSN: 0093-9994.
[5] Y. Li, Y. Gu, and T. C. Green, "Revisiting grid-forming and grid-following inverters: A duality theory," *IEEE Transactions on Power Systems*, vol. 37, no. 6, pp. 4541–4554, 2022.
[6] D. Dong, B. Wen, D. Boroyevich, P. Mattavelli, and Y. Xue, "Analysis of phase-locked loop low-frequency stability in three-phase grid-connected power converters considering impedance interactions," *IEEE Transactions on Industrial Electronics*, vol. 62, no. 1, pp. 310–321, 2014, ISSN: 0278-0046.
[7] J. Chen, W. Si, M. Liu, and F. Milano, "On the impact of the grid on the synchronization stability of grid-following converters," *IEEE Transactions on Power Systems*, vol. 38, no. 5, pp. 4970–4973, 2023.
[8] M. G. Dozein, B. C. Pal, and P. Mancarella, "Dynamics of inverter-based resources in weak distribution grids," *IEEE Transactions on Power Systems*, vol. 37, no. 5, pp. 3682–3692, 2022.
[9] B. Wen, D. Dong, D. Boroyevich, R. Burgos, P. Mattavelli, and Z. Shen, "Impedance-based analysis of grid-synchronization stability for three-phase paralleled converters," *IEEE Transactions on Power Electronics*, vol. 31, no. 1, pp. 26–38, 2015, ISSN: 0885-8993.
[10] K. M. Alawasa, Y. A.-R. I. Mohamed, and W. Xu, "Active mitigation of subsynchronous interactions between pwm voltage-source converters and power networks," *IEEE Transactions on Power Electronics*, vol. 29, no. 1, pp. 121–134, 2013, ISSN: 0885-8993.
[11] X. Zhang, S. Fu, W. Chen, N. Zhao, G. Wang, and D. Xu, "A symmetrical control method for grid-connected converters to suppress the frequency coupling under weak grid conditions," *IEEE Transactions on Power Electronics*, vol. 35, no. 12, pp. 13 488–13 499, 2020, ISSN: 0885-8993.
[12] J. A. Suul, S. D'Arco, P. Rodríguez, and M. Molinas, "Impedance-compensated grid synchronisation for extending the stability range of weak grids with voltage source converters," *IET Generation, Transmission & Distribution*, vol. 10, no. 6, pp. 1315–1326, 2016, ISSN: 1751-8687.
[13] P. S. Nguyen, G. Nourbakhsh, and G. Ledwich, "A novel virtual impedance method for interfacing renewable to grid with high impedance," in *2023 IEEE PES Innovative Smart Grid Technologies-Asia (ISGT Asia)*, IEEE, pp. 1–5, ISBN: 9798350327748.
[14] K. De Brabandere, T. Loix, K. Engelen, *et al.*, "Design and operation of a phase-locked loop with kalman estimator-based filter for single-phase applications," in *IECON 2006-32nd Annual Conference on IEEE Industrial Electronics*, IEEE, pp. 525–530, ISBN: 1509091556.
[15] H. Huang, Y. Lin, X. Lu, Y. Zhao, and A. Kumar, "Dynamic state estimation for inverter-based resources: A control-physics dual estimation framework," *IEEE Transactions on Power Systems*, 2024.
[16] H. Ahmed, S. Biricik, and M. Benbouzid, "Linear kalman filter-based grid synchronization technique: An alternative implementation," *IEEE Transactions on Industrial Informatics*, vol. 17, no. 6, pp. 3847–3856, 2020, ISSN: 1551-3203.
[17] M. Liserre, R. Teodorescu, and F. Blaabjerg, "Stability of photovoltaic and wind turbine grid-connected inverters for a large set of grid impedance values," *IEEE transactions on power electronics*, vol. 21, no. 1, pp. 263–272, 2006, ISSN: 0885-8993.
[18] J. Dannehl, M. Liserre, and F. W. Fuchs, "Filter-based active damping of voltage source converters with *LCL* filter," *IEEE Transactions on Industrial Electronics*, vol. 58, no. 8, pp. 3623–3633, 2010, ISSN: 0278-0046.
[19] T. V. Tran, S.-J. Yoon, and K.-H. Kim, "An lqr-based controller design for an lcl-filtered grid-connected inverter in discrete-time state-space under distorted grid environment," *Energies*, vol. 11, no. 8, p. 2062, 2018, ISSN: 1996-1073.
[20] R. Bimarta, T. V. Tran, and K.-H. Kim, "Frequency-adaptive current controller design based on lqr state feedback control for a grid-connected inverter under distorted grid," *Energies*, vol. 11, no. 10, p. 2674, 2018, ISSN: 1996-1073.
[21] J. Kukkola, M. Hinkkanen, and K. Zenger, "Observer-based state-space current controller for a grid converter equipped with an lcl filter: Analytical method for direct discrete-time design," *IEEE Transactions on Industry Applications*, vol. 51, no. 5, pp. 4079–4090, 2015, ISSN: 0093-9994.
[22] S.-J. Yoon, N. B. Lai, and K.-H. Kim, "A systematic controller design for a grid-connected inverter with lcl filter using a discrete-time integral state feedback control and state observer," *Energies*, vol. 11, no. 2, p. 437, 2018, ISSN: 1996-1073.
[23] L. Shieh, H. Wang, and R. Yates, "Discrete-continuous model conversion," *Applied Mathematical Modelling*, vol. 4, no. 6, pp. 449–455, 1980, ISSN: 0307-904X.
[24] B. D. Anderson and J. B. Moore, *Optimal filtering*. Courier Corporation, 2012, ISBN: 0486136892.
[25] B. D. Anderson and J. B. Moore, *Optimal control: linear quadratic methods*. Courier Corporation, 2007, ISBN: 0486457664.
[26] X. Wu, S. K. Panda, and J. Xu, "Analysis and control of the output instantaneous power for three phase pwm boost rectifier under unbalanced supply voltage conditions," in *IECON 2006-32nd Annual Conference on IEEE Industrial Electronics*, IEEE, 2006, pp. 1–6.
[27] S. G. Jorge, J. A. Solsona, C. A. Busada, L. C. Aguirre-Larrayoz, M. I. Martínez, and G. Tapia-Otaegui, "Non-linear control of the power injected into a weak grid by a self-synchronized inverter," *arXiv preprint arXiv:2410.07102*, 2024.
[28] S. G. Jorge, J. A. Solsona, C. A. Busada, G. Tapia-Otaegui, A. Susperregui, and M. I. Martínez, "Power control of converters connected via an l filter to a weak grid. a flatness-based approach," *arXiv preprint arXiv:2409.05527*, 2024.
[29] S. G. Jorge, J. A. Solsona, C. A. Busada, G. Tapia-Otaegui, A. S. Burguete, and M. I. M. Aguirre, "Nonlinear controller allowing the use of a small-size dc-link capacitor in grid-feeding converters," *IEEE Transactions on Industrial Electronics*, vol. 71, no. 3, pp. 2157–2166, 2023.
[30] J. A. Solsona, S. G. Jorge, and C. A. Busada, "A nonlinear control strategy for a grid-tie inverter that injects instantaneous complex power to the grid," in *2020 IEEE International Conference on Industrial Technology (ICIT)*, IEEE, 2020, pp. 895–900.
[31] G. Tapia-Otaegui, S. G. Jorge, J. A. Solsona, A. Susperregui, M. I. Martínez, and C. A. Busada, "Complex-variable sliding-mode control of instantaneous complex energy and power for grid-tied inverter," *IFAC-PapersOnLine*, vol. 56, no. 2, pp. 451–457, 2023.